\documentclass{jkas}

\def\year{2022} 
\def\month{February} 
\def\volume{00} 
\def\issue{0} 
\def\beginpage{1} 
\def\endpage{7} 
\def\received{mm dd, 2021} 
\def\accepted{mm dd, 2021} 
\date{Received \received; accepted \accepted}

\title{
V608 Cassiopeiae: A W UMa-type eclipsing binary with two possible circumbinary companions
}

\author[]{Jang-Ho Park}
\author[]{Jae Woo Lee}

\affil[]{Korea Astronomy and Space Science Institute, Daejeon 34055, Republic of Korea; \email{pooh107162@kasi.re.kr}}

\def\corrauthor{
J.-H. Park
}

\def\runningauthor{
Park \& Lee
}

\def\runningtitle{
V608 Cassiopeiae
}

\def\keywords{
\it
binaries: close --- binaries: eclipsing --- stars: individual (V608 Cassiopeiae) --- techniques: photometric
}

\def\abstracttext{
We present the photometric properties of V608 Cas from detailed studies of light curves and eclipse timings.
The light curve synthesis indicates that the eclipsing pair is an overcontact binary with parameters of $\Delta T$ = 155 K, $q$ = 0.328, and $f$ = 26 \%.
We detected the third light $\ell_{3}$, which corresponds to about 8 \% and 5 \% of the total systemic light in $V$ and $R$ bands, respectively.
Including our 6 timing measurements, a total of 38 times of minimum light were used for a period study.
It was found that the orbital period of V608 Cas has varied in some combination of an upward parabola and two periodic variations.
The continuous period increase with a rate of $+$3.99$\times$10$^{-7}$ d yr$^{-1}$
can be interpreted as a mass transfer from the secondary component to the primary star at a rate of 1.51$\times$10$^{-7}$ M$_\odot$ yr$^{-1}$.
The periods and semi-amplitudes of the two periodic variations are about $P_3$ = 16.0 yr and $P_4$ = 26.3 yr, and $K_3$ = 0.0341 d and $K_4$ = 0.0305 d, respectively.
The most likely explanation of both cycles is a pair of light-traveling time effects operated by
the possible presence of third and fourth components with estimated masses of $M_3$ = 2.20 M$_\odot$ and $M_4$ = 1.27 M$_\odot$ in eccentric orbits of $e_3$ = 0.66 and $e_4$ = 0.52.
Because the contribution of $\ell_{3}$ is very low compared to the estimated masses of two circumbinary objects, they can be inferred as very faint compact objects.
}

\begin{document}
\jkashead 



\section{Introduction \label{Sec1}}

W UMa-type eclipsing binaries are generally composed of two dwarfs with similar temperature and luminosity.
Both components are in contact by filling their Roche lobes, and have a common envelope.
The contact binaries have evolved through angular momentum loss by magnetic braking in a detached stage, and
it is thought that circumbinary object(s) may have played an important role physically and dynamically in this process \citep{bradstreet1994,pribulla2006}.
In reality, it can be the basis for supporting this scenario that the circumbinary objects have been found in many W UMa-type contact binary systems
\citep{d'angelo2006,pribulla2006,tokovinin2006,rucinski2007,eggleton 2008,eggleton 2009,rappaport2013}.
Photometrically, circumbinary objects can be found from the periodic changes of the times of minima in the eclipse timing diagram and the detection of a third light ($\ell_{3}$) in the light curves.
To present the specific characteristics such as the orbital elements, the estimated mass, and the light contribution, a precise analysis of observational data accumulated through long-term monitoring is required.

V608 Cas (GSC 4320-1035, 2MASS J02242609+7123066, Gaia DR2 545350624848618496) was discoverd by \citet{hubel1976} as a variable star with an orbital period of 0.47 days and an amplitude of about 1.00 mag.
The unfiltered light curve by \citet{blatter2001} showed the typical light variation of a W UMa-type binary with an eclipse depths of about 0.55 mag, and
the most suitable orbital period for their observational data was given as 0.38 days.
\citet{liu2016} estimated the fundamental parameters of V608 Cas through an analysis of their $VRI$ multi-band light curves.
They also found a continuous increase of the orbital period at a rate of $+$6.49$\times$10$^{-7}$ d yr$^{-1}$ from the eclipse timing $O$--$C$ diagram, and
the possibility of the presence of a sinusoidal variation was mentioned from the residuals of the quadratic term.
\citet{panpiboon2018} investigated the evolutionary status of this system through their absolute parameters.

The aim of this study is to present the physical properties of V608 Cas through our new binary modeling, and
to examine the characteristics of the orbital period change newly discovered through continuous monitoring of the times of minima.


\section{CCD Photometric Observations \label{Sec2}}

New CCD photometric observations of V608 Cas were made on ten nights from November 2018 to February 2021,
using an FLI 4K CCD camera and $VR$ filters attached to the 61-cm reflector at Sobaeksan Optical Astronomy Observatory (SOAO) in Korea.
The observations of the first season (2018) were carried out to secure complete multi-band light curves, and the others were conducted to collect additional eclipse timings.
Time-series CCD frames were processed with the IRAF/CCDRED package to correct for flat, bias, and dark images, and aperture photometry for the stars was accomplished with the IRAF/PHOT package.
Among the stars with a color index similar to V608 Cas in the observing field, GSC 4320-0549 and GSC 4319-1090,
which were stable without variations in brightness during our observing runs, were selected as comparison (C) and check stars (K), respectively.
The standard deviations of the (K$-$C) differences (1$\sigma$) are about $\pm$0.005 mag in both bandpasses.
As a result, we obtained a total of 411 individual points (208 in $V$, 203 in $R$) from the 2018 season observations and they are plotted in Figure \ref{Fig1}.


\section{Light curve Synthesis \label{Sec3}}

To obtain the photometric solutions of V608 Cas, we analyzed our $VR$ light curves.
The binary modeling was performed using the 2003 version of the Wilson-Devinney synthesis code \citep{wilson1971,van2003},
which adequately represents the geometric and photometric distortions of the components of a binary system
\footnote{More information is available at: \url{ftp://ftp.astro.ufl.edu/pub/wilson/lcdc2003/ebdoc2003.2feb2004.pdf.gz}}.
Among the modes supported by this code, mode 3 (contact binary) was adopted for synthesis.
Our analyses were separately performed in two cases:
the light curves were solved with basic parameters (hereafter Model 1), and were analyzed by including an additional component in basic parameters (hereafter Model 2).
This process is detailed in the next paragraph.
In the modeling, it was found that the less massive component was completely occulted by the heavier and larger component at the secondary minimum.
Therefore, we refer to the more massive component as the primary star for this system and fixed its temperature to be $T_{1}$ = 5222 $\pm$ 324 K taken from the {\it Gaia} DR2 \citep{gaia2018},
where the error indicates the typical uncertainty of this catalogue.
The gravity-darkening exponents ($g$) and the bolometric albedos ($A$) were adopted to be 0.32 and 0.5, respectively \citep{lucy1967,rucinski1969},
because they should have convective atmospheres, according to the temperature of both components presented in Table \ref{Tab1}.
The linear bolometric ($X$, $Y$) and monochromatic ($x$, $y$) limb-darkening coefficients were initialized from the values of \citet{van1993}.
A reliable mass ratio ($q$) of V608 Cas through spectroscopic observations has not yet been reported.
Therefore, we used the $q$-search method to find a suitable photometric mass ratio of the system.
In this process, the orbital inclination ($i$), the temperature of the secondary star ($T_2$), the dimensionless surface equipotential ($\Omega_1$ = $\Omega_2$), and
the monochromatic luminosity of the primary star ($L_1$) were considered as adjustment parameters.
The $q$-search result is shown in Figure \ref{Fig2}, where the arrow indicates the global minimum of the weighted sum of squared residuals ($\Sigma$) achieved at $q$ = 0.31.
We included the $q$ value as a adjustable parameter in subsequent analyses.

\begin{figure*}[!ht]
\centering
\includegraphics{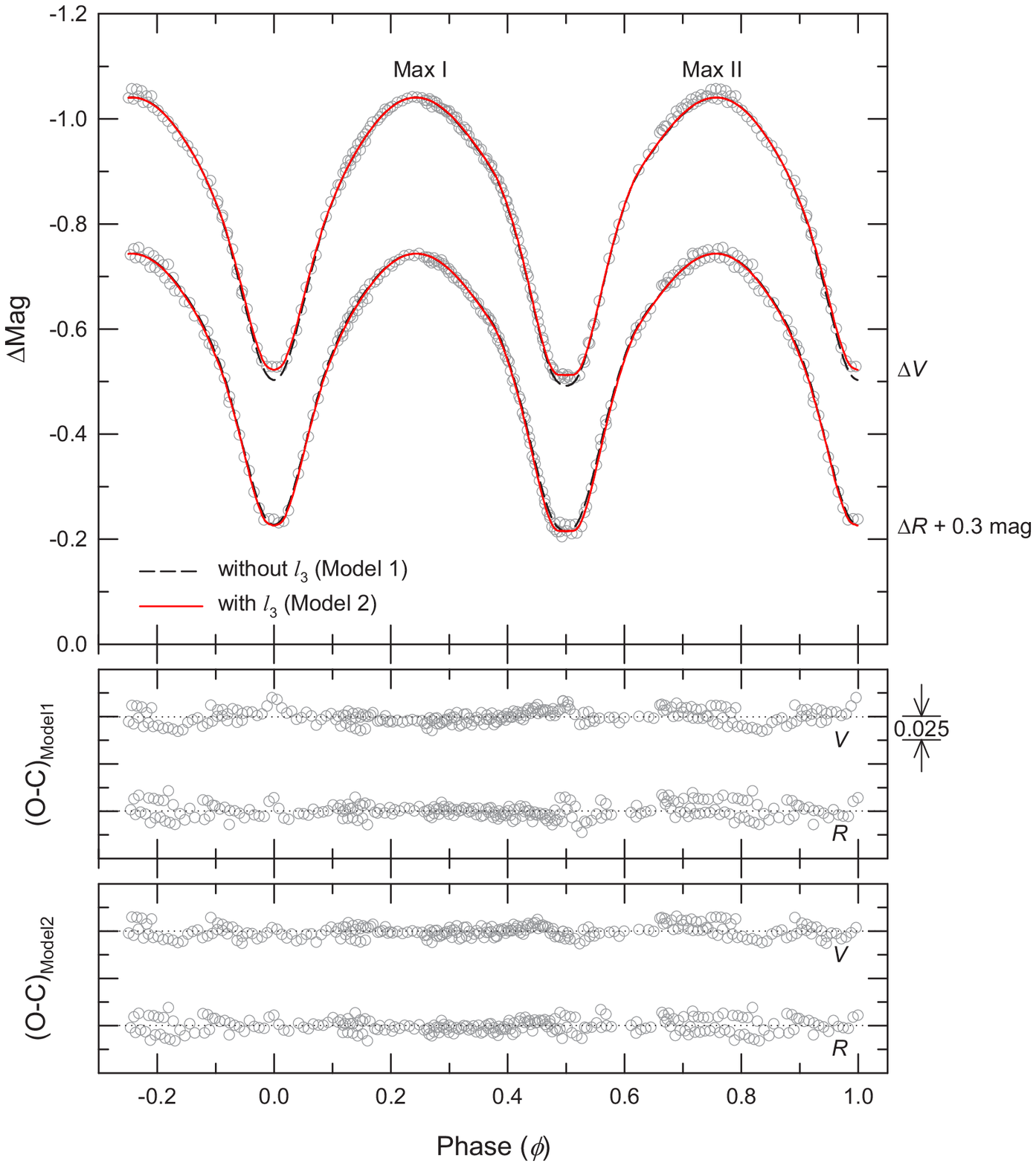}
\caption{Light curves of V608 Cas in $VR$ bandpasses as defined by individual observations.
The dashed and solid curves were computed without and with $\ell_{3}$ adjustment, respectively.
The middle and bottom panels show the light residuals from the solutions of Model 1 and Model 2 listed in Table \ref{Tab1}, respectively.}
\label{Fig1}
\end{figure*}

\begin{figure*}[!ht]
\centering
\includegraphics{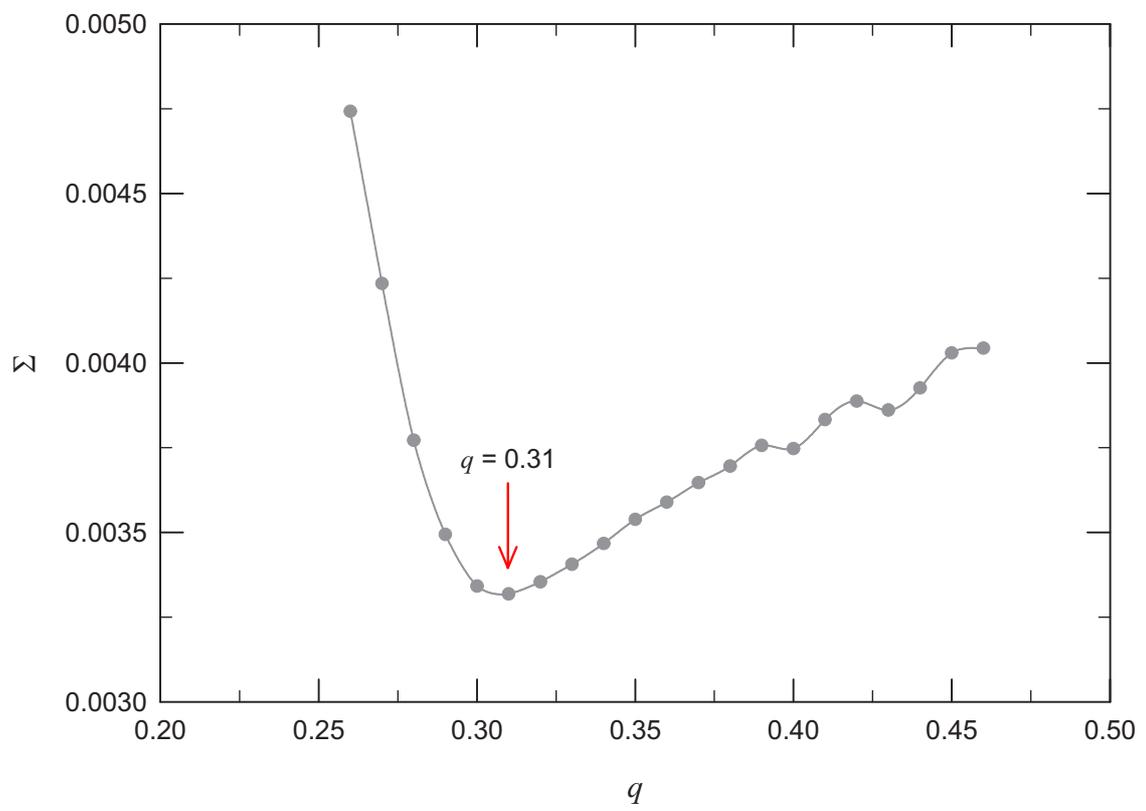}
\caption{Behavior of $\Sigma$ (the weighted sum of squared residuals) as a function of mass ratio $q$, indicating a minimum value at $q$ = 0.31 for V608 Cas.}
\label{Fig2}
\end{figure*}

The light curves of \citet{liu2016} and \citet{panpiboon2018} displayed the O'Connell effect in which Max II ($\phi$ = 0.75) was brighter than Max I ($\phi$ = 0.25), presumably due to dark spots on the massive primary star.
However, as shown in Figure \ref{Fig1}, ours did not show difference between Max I and Max II.
First, we analyzed the light curves with only basic parameters of $i$, $T_2$, $\Omega$, $q$, and $L_1$.
The solution is listed in the Model 1 column of Table \ref{Tab1}, and are presented as dashed curves in the top panel of Figure \ref{Fig1}.
The corresponding light residuals are plotted in the middle panel.
As shown in this figure, the model light curves do not fit well both of the eclipse minima, especially $V$ light curve.
In this case, the discordance of the eclipse depths may be caused by either a third light ($\ell_{3}$) dependent on the bandpasses
(e.g. \citealt{ogloza1998} for SW Lyn; \citealt{lee2006} for XZ Leo; \citealt{lee2014} for V407 Peg; \citealt{tian2019} for V723 Per), or
time-varying spot activity on the components (e.g. \citealt{kang2002} for TY UMa; \citealt{qian2005} for FG Hya).
From the eclipse timing analyses, we detected the possible presence of circumbinary objects around V608 Cas (see Section \ref{Sec4}).
Thus, the light curves were analyzed by including the $\ell_{3}$ in basic parameters ($i$, $T_2$, $\Omega$, $q$, and $L_1$).
The results are given in the Model 2 column of Table \ref{Tab1}.
The synthetic solid curves are presented in the top panel of Figure \ref{Fig1}, and the light residuals from this solution are plotted in the bottom panel.
From these results, we can see that Model 2 fits the observations better than Model 1 and gives a smaller value of $\Sigma$.
On the other hand, we could not obtain satisfactory results for our light curves with the applicable spot models.
Thus, we adopted the Model 2 as an optimal model for the contact binary system.

Our photometric solutions indicate that V608 Cas is an overcontact binary with parameters of $i$ = 80$^\circ$.7, temperature difference ($\Delta T$) = 155 K, $q$ = 0.328, and fill-out factor ($f$) = 26 \%.
The $\ell_{3}$ values contribute to 7.9 \% in $V$, and 4.9 \% in $R$ of the total systemic light.
The absolute parameters for each component were calculated by applying the parameters of Model 2 to the equation of \citet{mardirossian1980}.
The mass of the primary star was estimated from the \citet{pecaut2013} relation between temperature and mass.
The results are listed in the lower part of Table \ref{Tab1}.


\section{Orbital Period Study \label{Sec4}}

From our observations, a total of six new times of minimum light were determined with the method of \citet{kwee1956}.
Then, 32 timings were collected from \citet{kreiner2001} database and recent literature.
These timings are listed in Table \ref{Tab2}.
For the period analysis, the weights for the timings were derived from the inverse squares of the standard deviations ($\pm$0.0007 d) of the full $O$--$C$ residuals.

\begin{figure*}[!ht]
\centering
\includegraphics{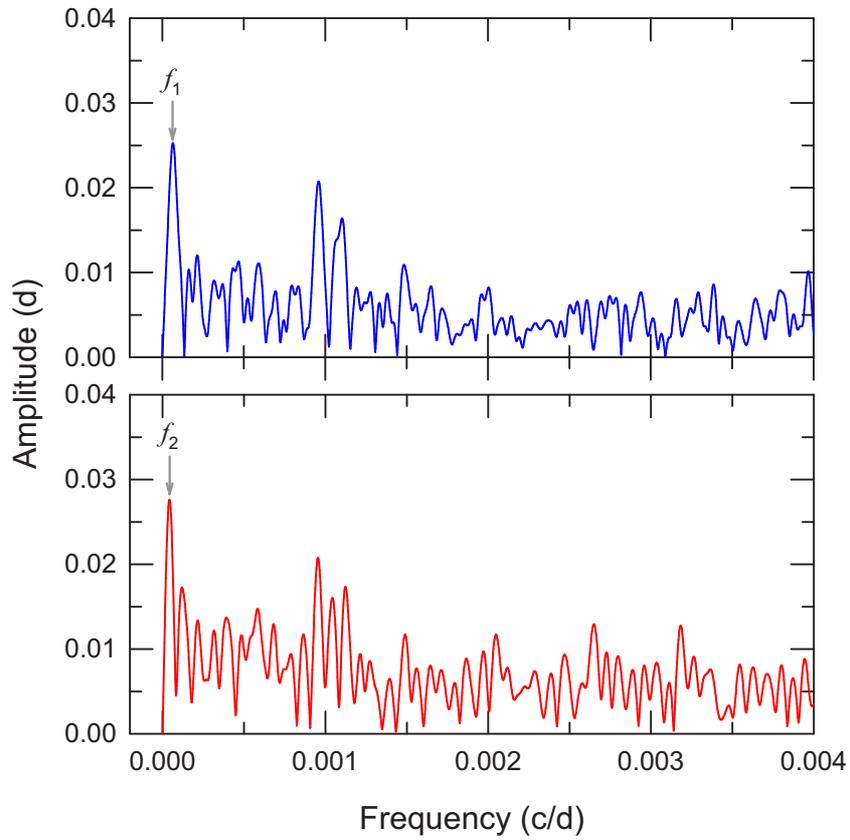}
\caption{Periodogram from the PERIOD04 formalism for the timing residuals from the quadratic ephemeris.
Two frequencies of $f_1$ = 0.0000650 cycle d$^{-1}$ and $f_2$ = 0.0000426 cycle d$^{-1}$ were detected by a pre-whitening process, as indicated by the arrows in the upper and lower panels.
These frequencies in turn correspond to periods of 16.0 yr and 24.4 yr, respectively.}
\label{Fig3}
\end{figure*}

The orbital period of V608 Cas shows a tendency to increase continuously in the form of an upward parabola.
A parabolic variation can be represented as $C = T_0 + PE + AE^2$ by adding a quadratic term ($AE^2$) to a linear ephemeris ($T_0 + PE$).
Therefore, we fitted all times of minimum light to this quadratic ephemeris.
Interestingly, sine-like variations were seen in the timing residuals.
Using the PERIOD04 \citep{lenz2005} program, which can extract individual frequencies from the multi-periodic content of an astronomical time series containing gaps,
we checked whether the residuals actually represent periodic variations.
As a result, as shown in Figure \ref{Fig3}, the two dominant frequencies were sequentially detected at $f_1$ = 0.0000650 cycle d$^{-1}$ and $f_2$ = 0.0000426 cycle d$^{-1}$ corresponding to about 16.0 yr and 24.4 yr, respectively.
The periodic variations suggest light-traveling time (LTT) effects operated by the presence of additional components orbiting the eclipsing pair.
Thus, the timing data were fitted to the following quadratic {\it plus} two-LTT ephemeris:
\begin{eqnarray}
C = T_0 + PE + AE^2 + \tau_{3} + \tau_{4},
\end{eqnarray}
where $\tau_{3}$ and $\tau_{4}$ are the LTT due to circumbinary objects in the system \citep{irwin1952,irwin1959}, and each include five parameters ($a_{12}\sin i_{3,4}$, $e$, $\omega$, $n$ and $T$).
The thirteen parameters of the ephemeris were calculated by applying the Levenberg-Marquart method \citep{press1992}, and summarized in Table \ref{Tab3}, including related quantities.
These and subsequent calculations have used our absolute parameters given in Table \ref{Tab1}.
The $O$--$C$ diagram of V608 Cas is plotted in the top panel of Figure \ref{Fig4}, where the dashed and solid curves represent the quadratic term and the full contribution, respectively.
The two panels in the middle represent the LTT orbits of $\tau_{3}$ and $\tau_{4}$, respectively.
The bottom panel shows the residuals for the complete ephemeris, and these are listed as $O$--$C_{\rm full}$ in the fourth column of Table \ref{Tab2}.
As seen in this figure, the quadratic {\it plus} two-LTT ephemeris gives us satisfactory representation for the eclipse timing variation.
The LTT orbits have periods of $P_3$ = 16.0 yr and $P_4$ = 26.3 yr, semi-amplitudes of $K_3$ = 0.0341 d and $K_4$ = 0.0305 d, and eccentricities of $e_3$ = 0.66 and $e_4$ = 0.52, respectively.
The mass functions of the circumbinary objects become $f(m_{3})$ = 0.934 M$_\odot$ and $f(m_{4})$ = 0.342 M$_\odot$, and
their estimated minimum masses are $M_3$ = 2.20 M$_\odot$ and $M_4$ = 1.27 M$_\odot$, respectively.

The quadratic term ($A$) in equation (1) denotes a continuous period increase of $+$3.99$\times$10$^{-7}$ d yr$^{-1}$,
which is a rate corresponding to a fractional period change of $+$1.09$\times$10$^{-9}$.
As the most common explanation, a period increase can be interpreted as a mass transfer from the less massive secondary component to the primary star,
in which the transfer rate is calculated as 1.51$\times$10$^{-7}$ M$_\odot$ yr$^{-1}$.


\section{Summary and Discussion}

In this study, we presented the photometric properties of V608 Cas, through detailed analyses of our new $VR$ light curves and the eclipse timing $O$--$C$ diagram.
We summarized our results as follows:

\begin{enumerate}
\item Our light curve synthesis shows that V608 Cas is an overcontact binary with a temperature difference of 155 K and a mass ratio of 0.328, and in a shallow contact of 26 \%.
The absolute parameters were estimated to be $M_1$ = 0.88 M$_\odot$ and $M_2$ = 0.29 M$_\odot$, $R_1$ = 1.15 R$_\odot$ and $R_2$ = 0.70 R$_\odot$, and $L_1$ = 0.88 L$_\odot$ and $L_2$ = 0.37 L$_\odot$, respectively.

\item The orbital period of V608 Cas has not varied in a monotonous way but through a combination of an upward parabola and
two periodic variations, with period lengths of $P_3$ = 16.0 yr and $P_4$ = 26.3 yr, and semi-amplitudes of $K_3$ = 0.0341 d and $K_4$ = 0.0305 d, respectively.
The continuous period increase of $+$3.99$\times$10$^{-7}$ d yr$^{-1}$
may be caused by the mass transfer of about 1.51$\times$10$^{-7}$ M$_\odot$ yr$^{-1}$ from the secondary component to the primary star.

\item The periodic variations can be interpreted as a pair of LTT effects due to the presence of two circumbinary objects around V608 Cas.
If the third and fourth companions are assumed to be the main sequence stars, their minimum masses are calculated as $M_3$ = 2.20 M$_\odot$ and $M_4$ = 1.27 M$_\odot$, respectively.
Because the $\ell_{3}$ contributions found from the light curves analysis are very low compared to the estimated masses,
each additional object could be a very faint compact object.
\end{enumerate}

\begin{figure*}[!ht]
\centering
\includegraphics{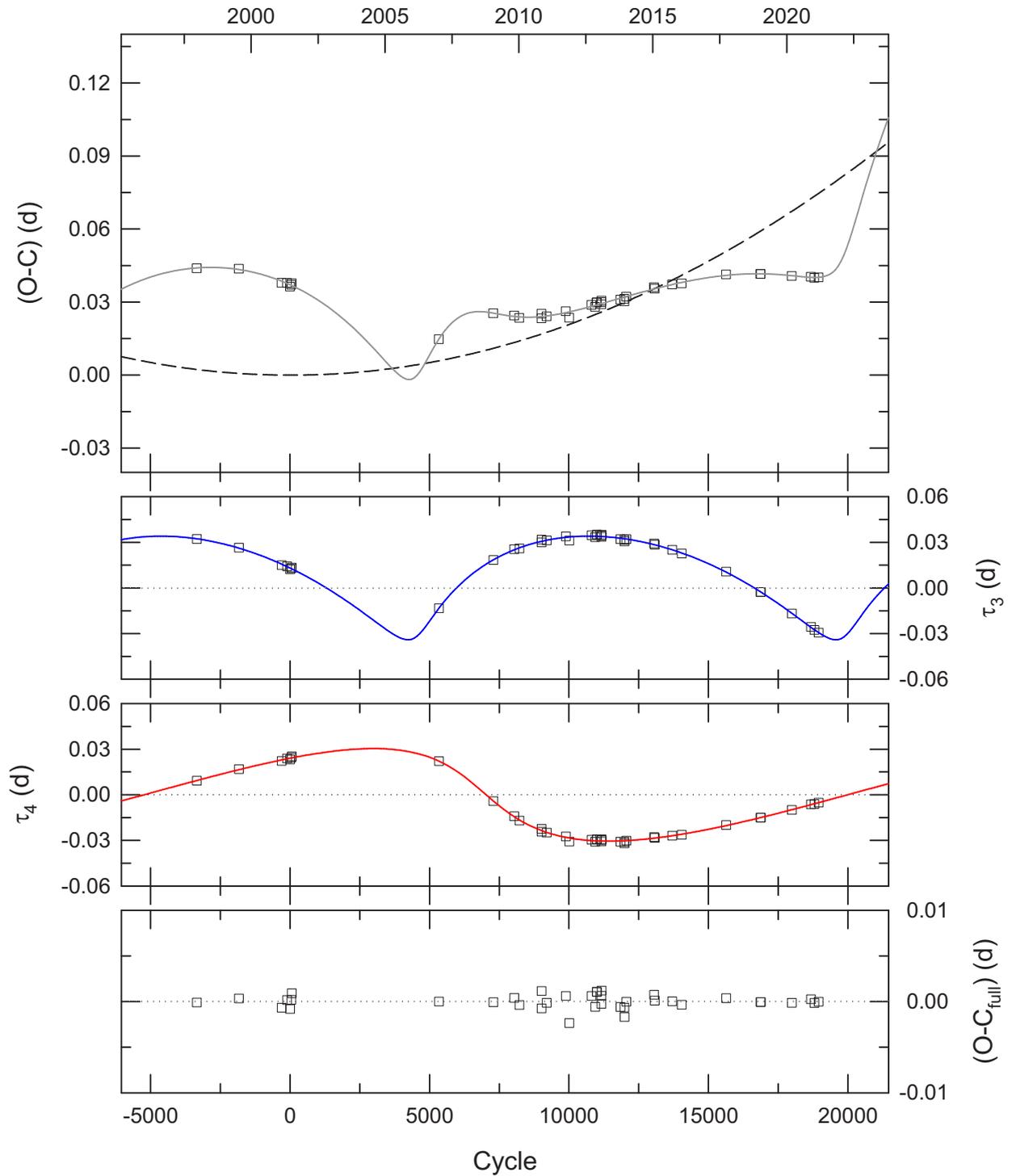}
\caption{Eclipse timing $O$--$C$ diagram of V608 Cas.
In the top panel, constructed with the linear terms of the quadratic {\it plus} two-LTT ephemeris, the dashed and solid curves represent the quadratic term and the full ephemeris, respectively.
The second and third panels represent the LTT orbits of $\tau_{3}$ and $\tau_{4}$, respectively.
The bottom panel shows the residuals from the complete ephemeris.}
\label{Fig4}
\end{figure*}

As shown in Figure \ref{Fig4}, all eclipse minima of V608 Cas coincide with our quadratic {\it plus} LTT ephemeris,
but there is a possibility of periodic deformations due to magnetic activity cycle of the components \citep{applegate1999,lanza1998}.
Therefore, we calculated the model parameters for each component by applying the periods ($P_{3,4}$) and amplitudes ($K_{3,4}$) listed in Table \ref{Tab3} to the Applegate formula.
The results are listed in Table \ref{Tab4}, where the rms luminosity changes ($\Delta m_{\rm rms}$) were converted to magnitude scale \citep{kim1997}.
Our results are that the variations of the gravitational quadrupole moment ($\Delta Q$) are two orders of magnitude smaller than typical values ($10^{51}\sim10^{52}$) for contact binaries \citep{lanza1999}, and
the rms luminosity variations ($\Delta L_{rms}$) represent that the Applegate mechanism cannot function in both components.
Consequentially, there is currently no other alternative to account for the sinusoidal variations than the LTT effects,
so they are most likely caused by two circumbinary objects around the eclipsing pair.
The putative circumbinary companions would have made an important contribution to the formation of the binary systems, and
these will evolve into a single star in the future through angular momentum loss.

We suggested the possibility of the circumbinary companions around V608 Cas.
The eclipse timing observations spanning 23 yr have covered only 1.4 cycle and 87 \% of the two periodic variations 16.0 yr and 26.3 yr.
Future high-precision timing measurements are important to identify the LTT effects and to understand their characteristics.


\acknowledgments
 
This paper was based on observations obtained at the Sobaeksan Optical Astronomy Observatory(SOAO),
which is operated by the Korea Astronomy and Space Science Institute (KASI).
We would like to thank the SOAO staff for assistance during our observations.
We also thanks C.-H. Kim for providing us the times of minimum light for V608 Cas.
We have used the Simbad database maintained at CDS, Strasbourg, France.



\begin{table*}[!ht]
\centering
\caption{Binary parameters of V608 Cas.}
\begin{tabular}{lccccc}
\toprule
Parameter                                & \multicolumn{2}{c}{Model 1}                 && \multicolumn{2}{c}{Model 2}                \\ \cline{2-3} \cline{5-6}
                                         & Primary           & Secondary               && Primary           & Secondary              \\
\midrule
Photometric solutions:                                                                                                               \\
$T_0$ (HJD)                              & \multicolumn{2}{c}{2,457,989.49765(6)}      && \multicolumn{2}{c}{2,457,989.49765(5)}     \\
$P_{\rm orb}$ (day)                      & \multicolumn{2}{c}{0.38040268(5)}           && \multicolumn{2}{c}{0.38040268(4)}          \\
$q$                                      & \multicolumn{2}{c}{0.317(2)}                && \multicolumn{2}{c}{0.328(2)}               \\
$i$ (deg)                                & \multicolumn{2}{c}{78.5(1)}                 && \multicolumn{2}{c}{80.7(1)}                \\
$T$ (K)                                  & 5222(324)$\rm ^a$ & 5332(331)               && 5222(324)$\rm ^a$ & 5377(334)              \\
$\Omega$                                 & 2.462(5)          & 2.462(5)                && 2.475(3)          & 2.475(3)               \\
$L_1$/($L_{1}$+$L_{2}$){$_{V}$}          & 0.7138(8)         & 0.2862                  && 0.6966(9)         & 0.3034                 \\
$L_1$/($L_{1}$+$L_{2}$){$_{R}$}          & 0.7182(6)         & 0.2818                  && 0.7030(9)         & 0.2970                 \\
$l_{3}${$_{V}$}                          & \multicolumn{2}{c}{\dots}                   && \multicolumn{2}{c}{0.079(5)}               \\
$l_{3}${$_{R}$}                          & \multicolumn{2}{c}{\dots}                   && \multicolumn{2}{c}{0.049(5)}               \\
$r$ (pole)$\rm ^b$                       & 0.4600(12)        & 0.2742(17)              && 0.4594(9)         & 0.2792(17)             \\
$r$ (side)$\rm ^b$                       & 0.4961(16)        & 0.2869(21)              && 0.4956(12)        & 0.2925(21)             \\
$r$ (back)$\rm ^b$                       & 0.5245(22)        & 0.3270(40)              && 0.5252(17)        & 0.3348(42)             \\
$r$ (volume)$\rm ^c$                     & 0.4928(17)        & 0.2952(26)              && 0.4927(13)        & 0.3013(26)             \\
$\sum W(O-C)^2$                          & \multicolumn{2}{c}{0.00329}                 && \multicolumn{2}{c}{0.00299}                \\
\midrule
Absolute parameters:                                                                                                                 \\
$M$ (M$_\odot$)                          & \multicolumn{2}{c}{\dots}                   && 0.88(9)           & 0.29(3)                \\
$R$ (R$_\odot$)                          & \multicolumn{2}{c}{\dots}                   && 1.15(4)           & 0.70(2)                \\
$L$ (L$_\odot$)                          & \multicolumn{2}{c}{\dots}                   && 0.88(27)          & 0.37(12)               \\
$a$ (R$_\odot$)                          & \multicolumn{2}{c}{\dots}                   && \multicolumn{2}{c}{2.33(8)}                \\
\bottomrule
\end{tabular}
\tabnote{$^{\rm a}$ Fixed parameter.}
\tabnote{$^{\rm b}$ Values of radii determined in each direction relative to semimajor axis ($a$).}
\tabnote{$^{\rm c}$ Mean volume radius calculated from the tables of \citet{mochnacki1984}.}
\label{Tab1}
\end{table*}

\begin{table*}[!ht]
\centering
\caption{CCD timings of the minimum light for V608 Cas.}
\begin{tabular}{lrrrcl}
\toprule
HJD           & Error         & Epoch           & $O$--$C_{\rm full}$ &  Min          &  References                         \\
(2,400,000+)  &               &                 &                     &               &                                     \\
\midrule
50,769.64     &               & $-$3343.5       &  $-$0.000109        &  II           & \citet{cook1999}                    \\
51,343.668    &               & $-$1834.5       &  $+$0.000336        &  II           & Paschke \citep{liu2016}             \\
51,926.4397   & $\pm$0.0009   &  $-$302.5       &  $-$0.000671        &  II           & \citet{blatter2001}                 \\
52,001.3790   & $\pm$0.0008   &  $-$105.5       &  $+$0.000167        &  II           & \citet{blatter2001}                 \\
52,041.5100   & $\pm$0.0008   &       0.0       &  $-$0.000820        &  I            & \citet{blatter2001}                 \\
52,058.4387   & $\pm$0.0007   &      44.5       &  $+$0.000178        &  II           & \citet{blatter2001}                 \\
52,065.4768   & $\pm$0.0009   &      63.0       &  $+$0.000919        &  I            & \citet{blatter2001}                 \\
54,071.3191   & $\pm$0.0024   &    5336.0       &  $+$0.000003        &  I            & \citet{hubscher2007}                \\
54,812.7353   & $\pm$0.0005   &    7285.0       &  $-$0.000062        &  I            & \citet{diethelm2009}                \\
55,100.8897   & $\pm$0.0006   &    8042.5       &  $+$0.000406        &  II           & \citet{diethelm2010}                \\
55,170.6927   & $\pm$0.0002   &    8226.0       &  $-$0.000370        &  I            & \citet{nelson2010}                  \\
55,473.3031   & $\pm$0.0016   &    9021.5       &  $-$0.000745        &  II           & \citet{hubscher2011}                \\
55,473.4952   & $\pm$0.0021   &    9022.0       &  $+$0.001153        &  I            & \citet{hubscher2011}                \\
55,542.7275   & $\pm$0.0002   &    9204.0       &  $-$0.000131        &  I            & \citet{diethelm2011}                \\
55,804.4469   & $\pm$0.0006   &    9892.0       &  $+$0.000618        &  I            & \citet{hubscher2012}                \\
55,850.8534   & $\pm$0.0006   &   10014.0       &  $-$0.002357        &  I            & \citet{diethelm2012}                \\
56,155.5615   & $\pm$0.0002   &   10815.0       &  $+$0.000606        &  I            & \citet{honkova2013}                 \\
56,203.8719   & $\pm$0.0003   &   10942.0       &  $-$0.000578        &  I            & \citet{diethelm2013}                \\
56,226.3175   & $\pm$0.0005   &   11001.0       &  $+$0.001055        &  I            & \citet{liu2016}                     \\
56,226.1273   & $\pm$0.0003   &   11000.5       &  $+$0.001059        &  II           & \citet{liu2016}                     \\
56,288.1322   & $\pm$0.0004   &   11163.5       &  $-$0.000259        &  II           & \citet{liu2016}                     \\
56,289.0841   & $\pm$0.0005   &   11166.0       &  $+$0.000625        &  I            & \citet{liu2016}                     \\
56,290.0357   & $\pm$0.0004   &   11168.5       &  $+$0.001210        &  II           & \citet{liu2016}                     \\
56,549.2808   & $\pm$0.0003   &   11850.0       &  $-$0.000590        &  I            & \citet{liu2016}                     \\
56,602.1572   & $\pm$0.0004   &   11989.0       &  $-$0.000667        &  I            & \citet{liu2016}                     \\
56,604.0582   &               &   11994.0       &  $-$0.001699        &  I            & \citet{liu2016}                     \\
56,629.1667   & $\pm$0.0002   &   12060.0       &  $-$0.000015        &  I            & \citet{liu2016}                     \\
57,007.6716   & $\pm$0.0002   &   13055.0       &  $+$0.000737        &  I            & \citet{nelson2015}                  \\
57,016.0399   & $\pm$0.0002   &   13077.0       &  $+$0.000104        &  I            & \citet{liu2016}                     \\
57,254.9347   & $\pm$0.0001   &   13705.0       &  $+$0.000029        &  I            & \citet{nelson2016}                  \\
57,383.5114   & $\pm$0.0001   &   14043.0       &  $-$0.000358        &  I            & \citet{hubscher2017}                \\
57,989.4971   & $\pm$0.0010   &   15636.0       &  $+$0.000373        &  I            & \citet{pagel2018}                   \\
58,460.05588  & $\pm$0.00008  &   16873.0       &  $-$0.000057        &  I            & SOAO (This Paper)                   \\
58,461.00690  & $\pm$0.00010  &   16875.5       &  $-$0.000044        &  II           & SOAO (This Paper)                   \\
58,884.96524  & $\pm$0.00007  &   17990.0       &  $-$0.000134        &  I            & SOAO (This Paper)                   \\
59,147.06264  & $\pm$0.00101  &   18679.0       &  $+$0.000252        &  I            & SOAO (This Paper)                   \\
59,192.13997  & $\pm$0.00018  &   18797.5       &  $-$0.000144        &  II           & SOAO (This Paper)                   \\
59,249.96141  & $\pm$0.00013  &   18949.5       &  $-$0.000020        &  II           & SOAO (This Paper)                   \\
\bottomrule
\end{tabular}
\label{Tab2}
\end{table*}

\begin{table*}[!ht]
\centering
\caption{Parameters for the quadratic {\it plus} two-LTT ephemeris of V608 Cas.}
\begin{tabular}{cccc}
\toprule
Parameter               &  \multicolumn{2}{c}{Quadratic {\it plus} Two-LTT}       &  Unit                  \\
\midrule
$T_0$                   &  \multicolumn{2}{c}{2,452,041.4737(2)}                  &  HJD                   \\
$P$                     &  \multicolumn{2}{c}{0.38040305(2)}                      &  day                   \\
$A$                     &  \multicolumn{2}{c}{2.08(1)$\times 10^{-10}$}           &  day                   \\
$a_{12}\sin i_{3,4}$    &  9.28(7)$\times 10^{8}$    &  9.25(8)$\times 10^{8}$    &  km                    \\
$e$                     &  0.66(1)                   &  0.52(1)                   &                        \\
$\omega$                &  297.8(6)                  &  895.8(5)                  &  deg                   \\
$n$                     &  0.06169(7)                &  0.03752(1)                &  deg d$^{-1}$          \\
$T$                     &  2,447,937(10)             &  2,416,274(15)             &  HJD                   \\
$P_{3,4}$               &  15.98(2)                  &  26.27(1)                  &  year                  \\
$K_{3,4}$               &  0.0341(3)                 &  0.0305(3)                 &  day                   \\
$f(m_{3,4})$            &  0.934(7)                  &  0.342(3)                  &  $M_\odot$             \\
$M_{3,4} \sin i_{3,4}$  &  2.195(14)                 &  1.267(8)                  &  $M_\odot$             \\
$dP$/$dt$               &  \multicolumn{2}{c}{3.99(2)$\times 10^{-7}$}            &  d yr$^{-1}$           \\
$dM$/$dt$               &  \multicolumn{2}{c}{1.51$\times 10^{-7}$}               &  $M_\odot$ yr$^{-1}$   \\
\bottomrule
\end{tabular}
\label{Tab3}
\end{table*}

\begin{table*}[!ht]
\centering
\caption{Applegate parameters for possible magnetic activities of V608 Cas.}
\begin{tabular}{ccccccc}
\toprule
Parameter                 & \multicolumn{2}{c}{Short-term ($\tau_{3}$)}       && \multicolumn{2}{c}{Long-term ($\tau_{4}$)}        &  Unit                \\ \cline{2-3} \cline{5-6}
                          & Primary                 & Secondary               && Primary                 & Secondary               &                      \\
\midrule
$\Delta P$                & \multicolumn{2}{c}{1.2050}                        && \multicolumn{2}{c}{0.6557}                        &  s                   \\
$\Delta P/P$              & \multicolumn{2}{c}{$3.67\times10^{-5}$}           && \multicolumn{2}{c}{$1.99\times10^{-5}$}           &                      \\
$\Delta Q$                & ${-1.87\times10^{50}}$  & ${-6.17\times10^{49}}$  && ${-1.02\times10^{50}}$  & ${-3.36\times10^{49}}$  &  g cm$^2$            \\
$\Delta J$                & ${6.70\times10^{47}}$   & ${3.23\times10^{47}}$   && ${3.65\times10^{47}}$   & ${1.76\times10^{47}}$   &  g cm$^{2}$ s$^{-1}$ \\
$I_s$                     & ${7.46\times10^{53}}$   & ${9.11\times10^{52}}$   && ${7.46\times10^{53}}$   & ${9.11\times10^{52}}$   &  g cm$^{2}$          \\
$\Delta \Omega$           & ${8.98\times10^{-7}}$   & ${3.54\times10^{-6}}$   && ${4.89\times10^{-7}}$   & ${1.93\times10^{-6}}$   &  s$^{-1}$            \\
$\Delta \Omega / \Omega$  & ${4.70\times10^{-3}}$   & ${1.85\times10^{-2}}$   && ${2.56\times10^{-3}}$   & ${1.01\times10^{-2}}$   &                      \\
$\Delta E$                & ${1.20\times10^{42}}$   & ${2.29\times10^{42}}$   && ${3.57\times10^{41}}$   & ${6.77\times10^{41}}$   &  erg                 \\
$\Delta L_{rms}$          & ${7.50\times10^{33}}$   & ${1.42\times10^{34}}$   && ${1.35\times10^{33}}$   & ${2.56\times10^{33}}$   &  erg s$^{-1}$        \\
                          & 1.95                    & 3.70                    && 0.35                    & 0.67                    &  L$_\odot$           \\
                          & 2.22                    & 10.01                   && 0.40                    & 1.80                    &  L$_{1,2}$           \\
                          & 1.02                    & 1.49                    && 0.27                    & 0.46                    &  mag                 \\
$B$                       & 22.6                    & 33.1                    && 13.0                    & 19.0                    &  kG                  \\
\bottomrule
\end{tabular}
\label{Tab4}
\end{table*}

\end{document}